\documentclass[12pt]{article}
\usepackage{epsfig}
\begin{document}

\centerline{\bf\Large{Determination of the dynamical parameters}}

\centerline{\bf\Large{of the Universe and its age}}
\vspace*{1.00 cm}

\centerline{Matts Roos \& S. M. Harun-or-Rashid} 
\centerline{Department of Physics, Division of High Energy Physics,} 
\centerline{University of Helsinki, Finland.} 

\vspace*{1.00 cm}

\centerline {\bf ABSTRACT} 

We determine the density parameters $\Omega_m$ of gravitating matter and
$\Omega_{\Lambda}$ of vacuum energy, by making a $\chi^2$ fit to nine
independent astrophysical constraints.  Paying rigorous attention to
statistical detail, we find that the present best values are
$\Omega_m=0.31\pm 0.07, \Omega_{\Lambda}=0.70\pm 0.13$, where these
$1\sigma$ errors are approximately Gaussian (thus trivially convertible
to whatever percentage confidence range desired).  The total $\chi^2$ is
2.5 for 7 degrees of freedom, testifying that the various systematic
errors included are generous.  Since $\Omega_m + \Omega_{\Lambda}=
1.01\pm 0.15$, it follows that the Einstein-de Sitter model is very
strongly ruled out, that also any low-density model with
$\Omega_{\Lambda}=0$ is ruled out, and that a flat cosmology is not only
possible, but clearly preferred.  In the flat case we find
$\Omega_m=0.31\pm 0.04$, from which it follows that the age of the
Universe is $t_0 = 13.7^{+1.2}_{-1.1}(0.68/h)$ Gyr. 

\section{INTRODUCTION}

If the dynamical parameters describing the cosmic expansion were known
to good precision, we would know whether the Universe is open or closed,
or whether its geo\-metry is in fact exactly flat as inflationary theory
wants it. To know the answer we need at least (i) the Hubble constant
$H_0$, usually given in the form $H_0 = 100h$ km s$^{-1}$ Mpc$^{-1}$,
(ii) the dimensionless density parameter $\Omega_m$ of gravitating
matter, comprising baryons, neutrinos and some yet unknown kinds of dark
matter, and (iii) the density parameter $\Omega_{\Lambda}$ of vacuum
energy, related to the cosmological constant $\Lambda$ by 
\begin{eqnarray}
\Omega_{\Lambda} = \Lambda/3H^2_0\ .
\end{eqnarray}         
A flat universe is defined by the condition
\begin{eqnarray}
\Omega_m + \Omega_{\Lambda}=1\ .
\label{f2}\end{eqnarray}

In a previous publication (Roos \& Harun-or-Rashid 1998) we tried to
determine the preferred region in the $(\Omega_m,
\Omega_{\Lambda})$-plane by combining three independent observational
constraints and a value for $H_0$.  We then arrived at the conclusion
that the standard Einstein-de Sitter model with $(\Omega_m,
\Omega_{\Lambda}) = (1, 0)$ was ruled out, that the preferred density
parameter ranges were (0.2-0.4, 0.8-0.6), and that the geometry of the
Universe therefore could be flat.  Since then several other constrained
fits with the same purpose have been published, however making use of
only a small number of constraints (some recent ones are Efstathiou et
al.  1998, Lahav \& Bridle 1998; Lineweaver 1998; Tegmark 1998; Tully
1998; Waga \& Miceli 1998; Webster 1998; White 1998).

The purpose of the present paper is to continue this pursuit of what
kind of universe we are living in. In this study we make use of nine
independent constraints of which only two are related to the constraints
we used before, but they are now based on much larger data samples. As
is certainly well known to the reader, the whole field is in a state of
rapid expansion both in the quantity of observations made and in the
quality of the results. As we shall see in the light of present data,
our previous conclusions are being upheld, now indeed with higher
confidence.

When $H_0, \Omega_m$ and $\Omega_{\Lambda}$ are known, the age of the
Universe, $t_0$, can be obtained from the Friedman-Lemaître model as
\begin{eqnarray}
t_0={{1}\over{H_0}}\int_0^1\hbox{d}x
\big{[}(1-\Omega_m-\Omega_{\Lambda})+\Omega_mx^{-1}+\Omega_{\Lambda}x^2
\big{]}^{-1/2}.  
\label{f3}\end{eqnarray}

In Section 2 we describe the nine observational constraints entering our
least-squares fit.  In Section 3 we describe the results of our fit to
these constraints, assuming that all reported observational errors as
well as systematic errors are Gaussian (unless explicitly stated
otherwise).  Paying rigorous attention to statistical detail, we present
our best value in the $(\Omega_m, \Omega_{\Lambda})$-plane as well as
along the flat line Eq.  (\ref{f2}).  We then also use Eq.  (\ref{f3})
to determine $t_0$ in the plane and on the line.  Section 4 contains a
discussion of the effects of systematic errors and a comparison with
related results not included in our fit.  Section 5 summarizes our
conclusions. 

\section{OBSERVATIONAL CONSTRAINTS}

\subsection{Cosmic Microwave Background Radiation}
The observations of anisotropies in the CMBR are commonly presented as
plots of the multipole moments $C_{\ell}$ against the multipole $\ell$,
or equivalently, against the FWHM value of the angular anisotropy
signal. This power spectrum has most recently been analysed by
Lineweaver (1998) and Tegmark (1998) in order to constrain possible
cosmological models. In general, the theoretical models for the power
spectrum may depend on up to 9 parameters. 

What interests us here is the confidence region in the marginal subspace
of the $(\Omega_m, \Omega_{\Lambda})$-plane.  The $1\sigma$ region then
appears to be approximately a wedge with straight-line edges and its
apex $A$ is at (0.15, 0.77).  The best fit point $O$ is located
asymmetrically within this wedge at (0.45, 0.35) (Fig. 2 in Lineweaver 1998). 
However, the $1\sigma$ line plotted in this graph corresponds to 
$\Delta\chi^2=1$ whereas it should be drawn at $\Delta\chi^2=2.3$ in a
two-dimensional parameter space.  Denoting the distance from $O$ to an
arbitrary point $P$ in the plane by $r$, and the distance from $O$
through $P$ to the $1\sigma$ wedge line by $r_0$, our constraint is
therefore of the form
\begin{eqnarray}
r^2/r_0^2 \ .
\label{f4}\end{eqnarray}
Fortunately the position of the point $O$ with respect to the $1\sigma$
wedge lines is so remote from the range preferred by the collection of
all constraints, that we do not have to worry about the asymmetry. 

\subsection{Gas fraction in X-ray clusters}

Matter in an idealized, spherically symmetric cluster is taken to be
distributed in two characteristic regions --- a nearly hydrostatic inner
body surrounded by an outer, infalling envelope. Outside this {\it
virial radius} separating these regions, all matter is infalling with
the cosmic mix of the components. A common definition of the virial
radius is $R_{500}$, the radius outside which the density drops below
500 in units of the critical density (Navarro, Frenk \& White 1995)  The
baryonic component of the mass in galaxy clusters is dominated by gas
which can be observed by its X-ray emission. Thus by measuring the gas
fraction near the virial radius one expects to obtain fairly unbiased
information on the ratio of $\Omega_m$ to the cosmic baryonic density
parameter $\Omega_b$.  

For this purpose Evrard (1997) has used a very large sample of clusters:
the ROSAT compilation of David, Jones \& Forman (1995) and the Einstein
compilation of White \& Fabian (1995). He has obtained a 'realistic'
value of 
\begin{eqnarray}
{\Omega_m\over \Omega_b} h^{-4/3} \approx (11.8\pm 0.7)\ .
\label{f4a}\end{eqnarray}
 This value includes a galaxy mass estimate of 20\% of gas mass, and a
baryon diminution $\Upsilon(500)=0.85$ at $R_{500}$.

Taking $\Omega_b=0.024\pm 0.006 h^{-2}$ from the low primordial
deuterium abundance (Tytler, Fan \& Burles 1996), and $h=0.68\pm 0.05$
from the analysis by Nevalainen \& Roos (1998) who studied the 
metallicity effect of Cepheid-calibrated galaxy distances, one obtains 
\begin{eqnarray}
\Omega_m=0.36\pm 0.09\ , 
\label{f5}\end{eqnarray}
which we use as our constraint.  Note that the lower limit of $\Omega_m$
only holds if dark matter is alltogether non-baryonic.
We shall come back to this problem in the Discussion.

\subsection{Cluster mass function and the Ly$\alpha$ forest}

In theories of structure formation based on gravitational instability
and Gaussian initial fluctuations, massive galaxy clusters can form
either by the collapse of large volumes in a low density universe, or by
the collapse of smaller volumes in a high density universe. This is
expressed by the cluster mass function which constrains a combination
of  $\Omega_m$ and the amplitude $\sigma$ of mass fluctuations
(normalized inside some volume). The amplitude $\sigma$ is given by an
integral over the mass power spectrum. 

Weinberg et al. (1998) estimate  $\Omega_m$ by combining the cluster
mass function constraint with the linear mass power spectrum  determined
from Ly$\alpha$ data. For $\Omega_{\Lambda}=0 $ they obtain
$\Omega_m=0.46^{+ 0.12}_{-0.10}$ and for a flat universe they obtain
$\Omega_m=0.34^{+ 0.13}_{-0.09}$. In the region of interest of our fits,
these results can be approximated by the relation
\begin{eqnarray}
\Omega_m +  0. 18 \Omega_{\Lambda}=0.46\pm 0.08
\label{f6}\end{eqnarray}
which we use as one constraint.

\subsection{X-ray cluster evolution}

Clusters of galaxies are the largest known gravitationally bound
structures in the Universe. Since they are thought to be formed by
contraction from density fluctuations in an initially fairly homogeneous
Universe, their distribution in redshift and their density spectrum as
seen in their X-ray emission gives precious information about their
formation and evolution with time.  Thus by combining the evolution in
abundance of X-ray clusters with their luminosity-temperature
correlation, one obtains a powerful test of the mean density of the
Universe,  $\Omega_m$. 

The results of  Bahcall, Fan \& Cen (1997) defines a $1\sigma$ band
intersecting $\Omega_{\Lambda}=0$ at $\Omega_m=0.3\pm 0.1$, and  
intersecting $\Omega_{\Lambda}=1-\Omega_m$  at $\Omega_m=0.34\pm 0.13$.
This result can be summarized in the relation 
\begin{eqnarray}
\Omega_m =  0.195\pm 0.11 + 0.071 \Omega_{\Lambda}
\label{f7}\end{eqnarray}
which we use as one constraint.

The results of  Eke et al. (1998) defines a $1\sigma$ band intersecting
$\Omega_{\Lambda}=0 $ at $\Omega_m=0.44\pm 0.2$, and  intersecting
$\Omega_{\Lambda}=1-\Omega_m$ at $\Omega_m=0.38\pm 0.2$. This analysis
uses data independent of  Bahcall, Fan \& Cen (1997) [we  do not use
Henry (1997) who analyses essentially the same data as Eke et al.
(1998)]. The above result can be summarized in the relation 
\begin{eqnarray}
\Omega_m =  0.44\pm 0.20 - 0.077 \Omega_{\Lambda}
\label{f8}\end{eqnarray}
which we use as one constraint.

\subsection{Gravitational lensing}

Chiba \& Yoshii (1999) have presented new calculations of gravitational
lens statistics in view of the recently revised knowledge of the
luminosity functions of elliptical and lenticular galaxies and their
internal dynamics.  They apply their revised lens model to a sample of
867 un\-dupli\-cated QSOs at $z > 1$ taken from several optical lens
surveys, as well as to 10 radio lenses. In sharp contrast to previous
models of lensing statistics that have supported a high-density universe
with  $\Omega_m=1$ , they conclude that a flat universe with 
$\Omega_m=0.3^{+0.2}_{-0.1}$ casts the best case to explain the results
of the observed lens surveys.

The number of multiply imaged QSOs found in lens surveys is sensitive
to  $\Omega_{\Lambda}$. Models of gravitational lensing must, however,
explain not only the observed probability of lensing, but also the
relative probability of showing a specific image separation. The image
separation increases with increasing $\sigma^*$, the characteristic
velocity dispersion. Thus the results can be expressed as likelihood
contour plots in the two-dimensional parameter space of $\sigma^*$ and
$\Omega_m=1-\Omega_{\Lambda}$. 

Instead of using the above quoted $\Omega_m$ value, we use the 68\%
likelihood contour in the two-dimensional parameter space of Fig. 8 of
Chiba \& Yoshii (1999), we integrate out $\sigma^*$, and we thus obtain
the one-dimensional 68\% confidence range 
\begin{eqnarray}
\Omega_{\Lambda}=0.70\pm 0.16\ .
\label{f9}\end{eqnarray}

\subsection{Classical double radio sources}

There are two independent measures of the average size of a radio
source, where size implies the separation of two hot spots: the average
size of similar sources at the same redshift, and the product of the
average rate of growth of the source and the total time for which the
highly collimated outflows of that source are powered by the AGN. This
outflow leads to the large scale radio emission. The two measures depend
on the angular size distance to the source in different ways, so
equating them allows a determination of the coordinate distance to the
source which, in turn, can be used to determine pairs of $\Omega_m,
\Omega_{\Lambda}$-values.  

Daly, Guerra \& Lin Wan (1998) have used 14 classical double radio
galaxies with redshifts $z\le 2$ to determine an approximately
elliptical 68\% confidence region in the ($\Omega_m,
\Omega_{\Lambda}$)-plane centered at (0.05, 0.32). In our fit this
constraint is represented by a term in the $\chi^2$-sum of the form
\begin{eqnarray}
[w^2 + z^2 -(\sigma^2_w -\sigma^2_z) (w^2/\sigma^2_w)]/\sigma^2_z\ ,
\label{f10}\end{eqnarray}
where $w$ and $z$ are the rotated coordinates
\begin{eqnarray}
w &= (\Omega_m - \Omega_{m,0}) \cos\theta + (\Omega_{\Lambda} -
\Omega_{\Lambda,0})\sin\theta\nonumber\\
z &= -(\Omega_m - \Omega_{m,0}) \sin\theta + (\Omega_{\Lambda} -
\Omega_{\Lambda,0})\cos\theta \ .
\label{f11}\end{eqnarray}
The rotation angle is $\theta = 70^{\circ}.4$ and  the $w$ and $z$
errors are $\sigma_w = 0.84, \sigma_z = 0.33$.

\subsection{Supernov\ae\ of type Ia}

Type Ia supernovae can be calibrated as standard candles, and have
enormous luminosities (at maximum, $\sim 10^{10} {\cal L}_\odot$). 
These two features make them a near-ideal tool for studying the
luminosity-redshift relationship at cosmological distances.  By
calibrated standard candles, we mean there is some correctable
dispersion among the magnitudes at light maximum.  The light curves are
very similar, but some are slightly wider and some narrower than
average, with width being correlated with brightness at maximum light. 
The wider, intrinsically brighter objects are also bluer and have minor
but recognizable spectral differences from the narrower, fainter, redder
objects. 

The factor relating brightness to redshift (as compared with 'nearby'
SNe Ia) is a function of $\Omega_m$ and $\Omega_\Lambda$.  If one has a
set of nearby and a set of distant supernovae all at about the same
redshift, then a nearly-linear relationship between $\Omega_m$ and
$\Omega_\Lambda$ can be found.  At a different redshift the slope is
different (Goobar \& Perlmutter 1995), and so for a set of SNe with
various redshifts, say between 0.5 and 1.0, one can find a best-fit
region in the $(\Omega_m,\Omega_\Lambda)$-plane.

The High-z Supernova Search Team (Riess et al.  1998) have used 10 SNe
Ia in the redshift range 0.16 -- 0.62 and 34 nearby supernov\ae\ to
place constraints on the dynamical parameters and $t_0$.  In the
($\Omega_m, \Omega_{\Lambda}$)-plane their 68\% confidence region is an
ellipse centered at (0.20, 0.65).  In our fit this constraint is
represented by a term in the $\chi^2$-sum of the same form as in Eq. 
(\ref{f10}) with the parameter values $\theta = 51^{\circ}.1$, $\sigma_w
= 1.27, \sigma_z = 0.18$. 

The Supernova Cosmology Project has so far published an analysis based
only on their first discovered six supernov\ae\ (Perlmutter et al.
1998). Since then the team has discovered many more supernov\ae\ , and a
preprint is now available describing the analysis and results of
42 supernov\ae\ in the high-redshift range 0.18 -- 0.83. The redshifts
have mainly been measured from the narrow host-galaxy lines, rather than
the broader supernova lines. The light curves are compared to a standard
'template' time function, which is then time-dilated by a factor $1+z$
to account for the cosmological lengthening of the time scale. The
conclusion in this paper that the Universe is today accelerating is very
strongly influencing our fit. However, we don't see it as our task to
invent possible systematic errors affecting this conclusion, we take the
results as they are presented.

In the ($\Omega_m, \Omega_{\Lambda}$)-plane the 68\% confidence range is
an ellipse centered at (0.75, 1.36). In our fit this constraint is
represented by a term in the $\chi^2$-sum of the same form as in Eq.
(\ref{f10}) with the parameter values $\theta = 54^{\circ}.35$, 
$\sigma_w = 1.3, \sigma_z = 0.15$.

\section{RESULTS}

We perform a least-squares fit to the above nine constraints using two
free parameters $\Omega_m, \Omega_{\Lambda}$.  The Hubble constant is
not treated as a free parameter at this stage, but is fixed to the
Nevalainen \& Roos (1998) value.  Actually it only enters when
constraining the gas fraction in X-ray clusters, and there the $H_0$
error is negligible compared to the errors specific to this constraint. 

For some of the constraints both a $1\sigma$ contour and a $2\sigma$
contour are given.  This constitutes useful information on the relevant
probability density function, which we take into account.  When only a
value and one error (of given confidence level) has been reported, we
treat the probability density function as if Gaussian.  When several
error terms have been reported for one measurement, e.g.  statistical
and systematic, we always add them quadratically. 

To find the absolute minimum in the ($\Omega_m, \Omega_{\Lambda}$)-plane
and the $1\sigma$ and $2\sigma$ contours (at $\Delta\chi^2=2.3$ and 6.2
up from the minimum, respectively, we use the standard minimization
program MINUIT (James \& Roos 1975). Although this program is widely
used in high energy physics, we digress briefly in Sec. 4.4 to describe it to
astronomers who might not know it.

The best fit value is then found to be \begin{eqnarray} \Omega_m=0.31\pm
0.07,\ \ \Omega_{\Lambda}=0.70\pm 0.13, \ \ \chi^2=2.5\ . 
\label{f12}\end{eqnarray} In Fig.1 we plot the shape of the $1\sigma$
and $2\sigma$ contours.  It is obvious from this figure that the errors
of the parameters are Gaussian to a very good precision. 

The above values can be added to yield
\begin{eqnarray}
\Omega_m + \Omega_{\Lambda}=1.01\pm 0.15\ .
\label{f12a}\end{eqnarray}

From this we conclude that (i) the data require a flat cosmology, (ii)
the Einstein-de Sitter model is very convincingly ruled out, and (iii)
also any low-density model with $\Omega_{\Lambda}=0$ is ruled out.  Note
that the results depend very strongly on the SNeIa data (Perlmutter et
al.  1998).  If we require exact flatness, the parameter values do not
change noticeably but the errors are of course smaller in this
one-dimensional space,
\begin{eqnarray} 
\Omega_m=0.31\pm 0.04 ,\ \ \Omega_{\Lambda}=0.69\pm
0.04,\ \ \chi^2=2.5\ .  
\label{f14}\end{eqnarray}

Let us now substitute the above parameter values and the accurate 
Hubble constant value $h=0.68\pm 0.05$ (Nevalainen \& Roos 1998) into
Eq. (\ref{f3}). [An equally accurate value, $h=0.69\pm 0.05$, has been
published by Giovanelli et al. (1997), obtained by using a Tully-Fisher
template relation with a kinematical zero point from the best 12
galaxies in a sample of 555 galaxies in 24 clusters.]  

Adding a 7.4\% $H_0$ error quadratically to the density parameter errors
in Eq.  (\ref{f12}) propagated through the integral (\ref{f3}), we find
as a value for the age of the Universe
\begin{eqnarray}
t_0 = 13.8\pm 1.3\ (0.68/h)\ \rm{Gyr}\ .
\label{f13}\end{eqnarray}
For an exactly flat Universe only the errors change slightly to
\begin{eqnarray}
t_0 = 13.7^{\ +1.2}_{\ -1.1}\ (0.68/h)\  \rm{Gyr}\ .
\label{f15}\end{eqnarray}

\section{DISCUSSION}

\subsection{Systematic errors}

With results as precise as those for the flat model, a question arising
is, what about neglected systematic errors? To this we have a
qualitative answer and a quantitative answer.

The constraints we use are indeed pulling the results in every
direction. CMBR is orthogonal to the supernova constraints, the
gravitational lensing constraint Eq. (\ref{f9}) is exactly orthogonal to
the gas fraction in X-ray clusters Eq. (\ref{f5}), and the remaining
constraints represent bands of several different directions.
Thus if many systematic errors have been neglected, their total effect
might largely be mutual cancellation.

This argument can be carried over to the systematic errors already
included: they represent pulls in every direction, so their distribution
must be quite random. Thus we think that it is justified to consider the
total systematic error to be random, just like the total statistical
error.

Let us now go to the quantitative answer. In the two-dimensional fit,
Eq. (\ref{f12}), $\chi^2$ is 2.5 for 7 degrees of freedom, and in the
one-dimensional fit, Eq. (\ref{f14}), $\chi^2$ is 2.5 for 8 degrees of
freedom. These $\chi^2$ values are much too low for statistically
distributed data. Thus we can conclude that the various errors quoted
for our nine constraints are not statistical: they have been blown up
unreasonably by the systematic errors added.

There are several reasons why this may have been done. One reason is
that several of these constraints have come from fits in parameter
spaces of higher dimension than just two. As a result, the quoted
$1\sigma$ confidence regions in the ($\Omega_m, \Omega_{\Lambda}$)-plane
which we use could be much too generous if they have not been properly
rescaled to lower dimensionality. 

Another reason is psychological.  When one adds systematic errors one
usually wants to have them big enough to play safe.  But that means that
one is really adding a 95\% CL (Confidence Level) or 99\% CL systematic
error to a 68\% CL statistical error.  The result is then a too large
total error, one not corresponding to a 68\% CL error. 

Note that under the assumption of a Gaussian probability density
function, a 99\% CL is not any 'safer' nor any more informative than a
68\% CL.  Both measures parametrize uniquely the width of one and the same
function, and they translate into one another by the mere multiplication
of a constant. 

Thus we conclude that our errors are already very generous, and that
there is no motivation for blowing them up with further arbitrary
systematic errors.

The most spectacular constraint is perhaps that of the Supernova Cosmology
Project (Perlmutter et al. 1998). Consequently one may worry about whether
unidentified systematic errors in those data might corrupt our conclusions.
To study that, we left out the Supernova Cosmology Project constraint, and
refitted the remaining 8 constraints. The result is then almost
unchanged from Eq. (\ref{f12}), 
\begin{eqnarray}
\Omega_m=0.30\pm 0.09,\ \ \Omega_{\Lambda}=0.70\pm 0.14, \ \ \chi^2=1.7\ ,
\label{f15a}\end{eqnarray}
and the sum of the parameters is
\begin{eqnarray}
\Omega_m + \Omega_{\Lambda}=0.99\pm 0.16\ .
\label{f15b}\end{eqnarray}
Thus our result is robust.

\subsection{Comparison with other data}

There are some categories of data which we have not used, but to which
it is nevertheless interesting to compare our results. Our selection
consists only of observations quoting a value and an error, but in
addition many interesting limits also exist.

Totani, Yoshii \& Sato (1997) have tested cosmological models for the
evolution of galaxies and star creation against the evolution of galaxy
luminosity densities. They have found $\Omega_{\Lambda}>0.53$
at 95\% confidence in a flat universe. This is inside our one-sided 95\%
confidence limit $\Omega_{\Lambda}>0.62$. 

Willick et al. (1997) have been comparing Tully-Fisher data for 838
galaxies within $cz\le 3000$ km/s from the Mark III catalog with the
peculiar velocity and density fields predicted from the 1.2 Jy IRAS
redshift survey. Taking $H_0$ in the range $55\le H_0\le 85$ (which is
three times too wide a range) and the spectral index $n$ of the rms
mass fluctuation power spectrum to be in the range 0.9--1.0, they
conclude that $0.16\le \Omega_m\le 0.40$, which covers our value in Eq.
(\ref{f14}). 

Danos \& Pen (1998) have determined an upper limit to $\Omega_m$ from
the apparent evolution of gas fractions in three rich clusters at
$z>0.5$. For the case of a flat universe they quote $\Omega_m<0.63$ at
95\% confidence. Our result from Eq. (\ref{f14}) is considerably more
stringent, $\Omega_m<0.38$ at 95\% confidence. 

Falco, Kochanek \& Munoz (1998) have determined the redshift
distribution of 124 radio sources and used it to derive a limit on
$\Omega_m$ from the statistics of six gravitational lenses. In a flat
universe their best fit yields $\Omega_m>0.26$ at 95.5\% confidence,
whereas our result (\ref{f14}) corresponds to the limit $\Omega_m>0.24$.

Im, Griffiths \& Ratnatunga (1997) use seven field elliptical galaxies
to determine $\Omega_{\Lambda}=0.64^{+0.15}_{-0.26}$, in good agreement
with Eq. (\ref{f14}). We do not use this result as a constraint in the
$\Omega_m, \Omega_{\Lambda}$-plane because it explicitly refers only to
the flat model.

The data on X-ray clusters show enormous scatter and the various
analyses published to date appear very model-dependent and mutually
contradictory, divided into papers with low or intermediate
$\Omega_m$-results and papers favoring the Einstein--de Sitter model
with $\Omega_m=1$. Above we made use of data from the lower set, quoting
values ruling out $\Omega_m=1$ by several standard deviations (Bahcall,
Fan \& Cen 1997; Henry 1997; Eke et al. 1998).The higher set we think is
less useful because it exhibits very shallow likelihood functions weakly
supporting almost any value in the range from 0.3 to 1.0 (within 90\%
CL) (Sadat, Blanchard \& Oukbir 1998; Reichart et al. 1998; Blanchard \&
Bartlett 1998; Viana \& Liddle 1998). An analysis of the possible
reasons for the contradictions and the effects of known systematics can
be found in Eke et al. (1998).

An upper limit to $\Omega_m$ can also be found by measuring the
Sunyaev-Zel'dovich effect in rich clusters, which gives directly a lower
limit to the total baryon fraction. Using the BBN prediction $\Omega_b
h^2=0.013\pm 0.002$ from the high deuterium abundance observation
(Rugers \& Hogan 1996), Myers et al. (1997) derive the upper limit
$\Omega_m h\le 0.21\pm 0.05$. Here we do not understand what the limit
means, because a limit cannot be a number with errors. If we take it to
mean $\Omega_m h\le 0.26$ at 68\% CL, our corresponding limit using
$h=0.68\pm 0.05$ is $\Omega_m h\le 0.24$, thus in good agreement.

However, since the deuterium abundance is such a sensitive and
contradictory issue, we would prefer to turn the logic of Myers et al.
(1997) around. Let us take the value of $\Omega_m$ from Eq. (\ref{f14}),
the value of $h$ from Nevalainen \& Roos (1998), then the
Sunyaev-Zel'dovich effect data can be used to derive a 68\% confidence
limit for $\Omega_b$, 
\begin{eqnarray}
\Omega_b h^2>0.015\ .
\label{f16}\end{eqnarray}

It may be interesting to go back to the Evrard (1997) analysis from
which we quoted a value in Eq. (\ref{f4a}) of Sec. 2.2. If instead, we
derive a value for the corresponding quantity using the low deuterium
abundance data of Tytler, Fan \& Burles (1996), the value of $\Omega_m$
from Eq. (\ref{f14}), and the value of $h$ from Nevalainen \& Roos
(1998), we find
\begin{eqnarray}
{\Omega_m\over \Omega_b} h^{-4/3} =10.0\pm 2.9\ .
\label{f17}\end{eqnarray}
The large error reflects the uncertainty in the deuterium abundance, but
given that we agree with Evrard's value in Eq. (\ref{f4a}), $11.8\pm
0.7$.

Since we also determine a value for the age of the Universe, it is of
interest to look at other recent $t_0$ determinations. Chaboyer (1998)
refers to recent work on nucleochronology for an estimate of the oldest
stars, $t_{stars}=15.2\pm 3.7$ Gyr, a value too inaccurate to be of much
interest, and to several techniques which taken together determine the
age of the oldest globular clusters (GC) to $t_{GC}=11.5\pm 1.3$ Gyr. 
Another estimate of the age of the oldest globular clusters is due to
Jimenez (1998). He quotes the 99\% confidence range $t_{GC}=13.25\pm
2.75$ Gyr, which translates into the 68\% confidence range
$t_{GC}=13.3\pm 1.1$ Gyr (in the abstract Jimenez (1998) calls this
$t_{GC}=13.5\pm 2$ Gyr which we do not use).  Thus there is some
disagreement of the order of 1.8 Gyr between Chaboyer (1998) and Jimenez
(1998), motivating us to take a mid-value and add a 0.9 Gyr systematic
error,
\begin{eqnarray}
t_{GC} = 12.4\pm 1.3\pm 0.9\ \rm{Gyr}\ .
\label{f18}\end{eqnarray}

In order to obtain $t_0$ from the GC value, however, one must add the
time it took for the metal-poor stars to form. Estimates are poor due to
the lack of a good theory, so Chaboyer (1998) recommends adding 0.1 to 2
Gyr. We presume that this is a 90\% CL estimate, $+1\pm 1$ Gyr. Thus one
arrives at the 68\% confidence range 
\begin{eqnarray}
t_0 = 13.4\pm 1.3\pm 0.9\pm 0.6\ \rm{Gyr} = 13.4\pm 1.7\ \rm{Gyr}\ .
\label{f19}\end{eqnarray}

Jimenez (1998) also finds an age of $t_0 = 13\pm 2$ Gyr for the red
elliptical galaxy 53W069 at $z=1.43$. This value as well as Eq.
(\ref{f19}) are in excellent agreement with the somewhat more accurate
Eq. (\ref{f15}).

\subsection{Cluster evolution and large-scale structure modelling}

In the subsection on X-ray cluster evolution we pointed out that
the data on X-ray clusters show enormous scatter and the various
analyses published to date are very model-dependent and mutually
contradictory. Although the authors of these studies defend the view
that this is a good method to determine $\Omega_m$, we think that the
data would be better used if the value of $\Omega_m$ would be taken from
elsewhere, e.g. from our Eq. (\ref{f14}), in order to learn more about
cluster evolution modelling.

The same comment applies to most of large-scale structure modelling. Not
long ago Monte Carlo simulations of existing models were still used to
defend the Einstein-de Sitter model, which was disfavored or ruled out
by almost every piece of information from other fields, and which we now
rule out by some 16 standard deviations. The understanding of galaxy
formation and cluster formation models would probably take a big step
forward if one used as a starting point the dynamic parameters derived
in this study.

\subsection{Numerical analysis by MINUIT}

MINUIT (James \& Roos 1975) is conceived as a tool to find the minimum
value of a multi-parameter function $F$({\bf X}) in the space of the
parameters {\bf X}, and to analyze the shape of the function around the
minimum.  The principal application is foreseen for statistical
analysis, working on $\chi^2$ or log-likelihood functions, to compute
the best-fit parameter values and uncertainties, including correlations
between the parameters.  The space may be limited by physical
restrictions on the allowed values of the parameters (constrained
minimization).  The function $F$({\bf X}) need not be known
analytically, but it is specified by giving its value at any point {\bf
X}.  Minimization proceeds by evaluating $F$({\bf X}) repeatedly at 
different points {\bf X} determined by the
minimization algorithm used (there is a choice between different algorithms), 
until some minimum is
attained (as defined by chosen convergence criteria).  At that point the
numerical covariance matrix found is used to give information on the
errors (of chosen confidence) of the parameters. MINUIT is available
from the CERN Program Library and is documented there as entry D506.

\section{SUMMARY}

With the purpose of determining the present best values of the density
parameters $\Omega_m$ of gravitating matter and the density parameter
$\Omega_{\Lambda}$ of vacuum energy, related to the cosmological
constant $\Lambda$, we have made a $\chi^2$ fit to nine independent
astrophysical constraints.  These are:\\
(i) the anisotropies in the CMBR as expressed in terms of multipole
moments (Lineweaver 1998; Tegmark 1998),\\ 
(ii) measurements of the gas fraction near the virial radius in X-ray
clusters (Evrard 1997),\\
(iii) combining the cluster mass function with the linear mass power
spectrum in Ly$\alpha$ data (Weinberg et al. 1998),\\ 
(iv-v) combining the evolution in abundance of X-ray clusters with their
luminosity-temperature correlation (Bahcall, Fan \& Cen 1997; Eke et al.
1998),\\
(vi) gravitational lensing (Chiba \& Yoshii 1998) ,\\
(vii) measurements of the average separation of hot spots in classical
double radio sources (Daly, Guerra \& Lin Wan 1998),\\
(viii-ix) measurements of the luminosity distance of supernov\ae\ of
type Ia via the magnitude-redshift relation, as reported by the High-z
Supernova Search Team (Riess et al. 1998) and the Supernova Cosmology
Project (Perlmutter et al. 1998). 

Assuming that all reported observational errors as well as systematic
errors are Gaussian (unless explicitly stated otherwise), and paying
rigorous attention to statistical detail, we find that these constraints
exhibit an agreement better than statistical, $\chi^2=2.5$ for 7 degrees
of freedom.  This indicates that the included systematic errors have
been overestimated rather than vice versa.  Our constrained fit finds
the best point in the parameter space to be at

\begin{eqnarray}
\Omega_m=0.31\pm 0.07,\ \ \Omega_{\Lambda}=0.70\pm 0.13\ .
\end{eqnarray}

\vspace{1.0cm}

\begin{center}
% GNUPLOT: LaTeX picture
\setlength{\unitlength}{0.240900pt}
\ifx\plotpoint\undefined\newsavebox{\plotpoint}\fi
\begin{picture}(1050,630)(0,0)
\font\gnuplot=cmr10 at 10pt
\gnuplot
\sbox{\plotpoint}{\rule[-0.200pt]{0.400pt}{0.400pt}}%
\put(120.0,82.0){\rule[-0.200pt]{4.818pt}{0.400pt}}
\put(100,82){\makebox(0,0)[r]{0.4}}
\put(969.0,82.0){\rule[-0.200pt]{4.818pt}{0.400pt}}
\put(120.0,167.0){\rule[-0.200pt]{4.818pt}{0.400pt}}
\put(100,167){\makebox(0,0)[r]{0.5}}
\put(969.0,167.0){\rule[-0.200pt]{4.818pt}{0.400pt}}
\put(120.0,251.0){\rule[-0.200pt]{4.818pt}{0.400pt}}
\put(100,251){\makebox(0,0)[r]{0.6}}
\put(969.0,251.0){\rule[-0.200pt]{4.818pt}{0.400pt}}
\put(120.0,336.0){\rule[-0.200pt]{4.818pt}{0.400pt}}
\put(100,336){\makebox(0,0)[r]{0.7}}
\put(969.0,336.0){\rule[-0.200pt]{4.818pt}{0.400pt}}
\put(120.0,421.0){\rule[-0.200pt]{4.818pt}{0.400pt}}
\put(100,421){\makebox(0,0)[r]{0.8}}
\put(969.0,421.0){\rule[-0.200pt]{4.818pt}{0.400pt}}
\put(120.0,505.0){\rule[-0.200pt]{4.818pt}{0.400pt}}
\put(100,505){\makebox(0,0)[r]{0.9}}
\put(969.0,505.0){\rule[-0.200pt]{4.818pt}{0.400pt}}
\put(120.0,590.0){\rule[-0.200pt]{4.818pt}{0.400pt}}
\put(100,590){\makebox(0,0)[r]{1}}
\put(969.0,590.0){\rule[-0.200pt]{4.818pt}{0.400pt}}
\put(120.0,82.0){\rule[-0.200pt]{0.400pt}{4.818pt}}
\put(120,41){\makebox(0,0){0}}
\put(120.0,570.0){\rule[-0.200pt]{0.400pt}{4.818pt}}
\put(244.0,82.0){\rule[-0.200pt]{0.400pt}{4.818pt}}
\put(244,41){\makebox(0,0){0.1}}
\put(244.0,570.0){\rule[-0.200pt]{0.400pt}{4.818pt}}
\put(368.0,82.0){\rule[-0.200pt]{0.400pt}{4.818pt}}
\put(368,41){\makebox(0,0){0.2}}
\put(368.0,570.0){\rule[-0.200pt]{0.400pt}{4.818pt}}
\put(492.0,82.0){\rule[-0.200pt]{0.400pt}{4.818pt}}
\put(492,41){\makebox(0,0){0.3}}
\put(492.0,570.0){\rule[-0.200pt]{0.400pt}{4.818pt}}
\put(617.0,82.0){\rule[-0.200pt]{0.400pt}{4.818pt}}
\put(617,41){\makebox(0,0){0.4}}
\put(617.0,570.0){\rule[-0.200pt]{0.400pt}{4.818pt}}
\put(741.0,82.0){\rule[-0.200pt]{0.400pt}{4.818pt}}
\put(741,41){\makebox(0,0){0.5}}
\put(741.0,570.0){\rule[-0.200pt]{0.400pt}{4.818pt}}
\put(865.0,82.0){\rule[-0.200pt]{0.400pt}{4.818pt}}
\put(865,41){\makebox(0,0){0.6}}
\put(865.0,570.0){\rule[-0.200pt]{0.400pt}{4.818pt}}
\put(989.0,82.0){\rule[-0.200pt]{0.400pt}{4.818pt}}
\put(989,41){\makebox(0,0){0.7}}
\put(989.0,570.0){\rule[-0.200pt]{0.400pt}{4.818pt}}
\put(120.0,82.0){\rule[-0.200pt]{209.342pt}{0.400pt}}
\put(989.0,82.0){\rule[-0.200pt]{0.400pt}{122.377pt}}
\put(120.0,590.0){\rule[-0.200pt]{209.342pt}{0.400pt}}
\put(964,-2){\makebox(0,0)[r]{$\Omega_m$}}
\put(33,548){\makebox(0,0)[r]{$\Omega_\Lambda$}}
\put(964,-2){\makebox(0,0)[r]{$\Omega_m$}}
\put(33,548){\makebox(0,0)[r]{$\Omega_\Lambda$}}
\put(120.0,82.0){\rule[-0.200pt]{0.400pt}{122.377pt}}
\sbox{\plotpoint}{\rule[-0.400pt]{0.800pt}{0.800pt}}%
\put(511,227){\usebox{\plotpoint}}
\put(511,226.84){\rule{4.818pt}{0.800pt}}
\multiput(511.00,225.34)(10.000,3.000){2}{\rule{2.409pt}{0.800pt}}
\put(531,230.34){\rule{2.600pt}{0.800pt}}
\multiput(531.00,228.34)(6.604,4.000){2}{\rule{1.300pt}{0.800pt}}
\multiput(543.00,235.40)(0.863,0.516){11}{\rule{1.533pt}{0.124pt}}
\multiput(543.00,232.34)(11.817,9.000){2}{\rule{0.767pt}{0.800pt}}
\multiput(558.00,244.40)(0.627,0.520){9}{\rule{1.200pt}{0.125pt}}
\multiput(558.00,241.34)(7.509,8.000){2}{\rule{0.600pt}{0.800pt}}
\multiput(568.00,252.40)(0.485,0.516){11}{\rule{1.000pt}{0.124pt}}
\multiput(568.00,249.34)(6.924,9.000){2}{\rule{0.500pt}{0.800pt}}
\multiput(578.40,260.00)(0.526,0.562){7}{\rule{0.127pt}{1.114pt}}
\multiput(575.34,260.00)(7.000,5.687){2}{\rule{0.800pt}{0.557pt}}
\multiput(585.40,268.00)(0.526,0.650){7}{\rule{0.127pt}{1.229pt}}
\multiput(582.34,268.00)(7.000,6.450){2}{\rule{0.800pt}{0.614pt}}
\put(590.34,277){\rule{0.800pt}{1.927pt}}
\multiput(589.34,277.00)(2.000,4.000){2}{\rule{0.800pt}{0.964pt}}
\put(592.34,285){\rule{0.800pt}{4.095pt}}
\multiput(591.34,285.00)(2.000,8.500){2}{\rule{0.800pt}{2.048pt}}
\put(594.34,311){\rule{0.800pt}{1.927pt}}
\multiput(593.34,311.00)(2.000,4.000){2}{\rule{0.800pt}{0.964pt}}
\put(595.84,319){\rule{0.800pt}{2.168pt}}
\multiput(595.34,319.00)(1.000,4.500){2}{\rule{0.800pt}{1.084pt}}
\put(595.0,302.0){\rule[-0.400pt]{0.800pt}{2.168pt}}
\put(595.84,336){\rule{0.800pt}{4.095pt}}
\multiput(596.34,336.00)(-1.000,8.500){2}{\rule{0.800pt}{2.048pt}}
\put(594.34,353){\rule{0.800pt}{1.927pt}}
\multiput(595.34,353.00)(-2.000,4.000){2}{\rule{0.800pt}{0.964pt}}
\put(592.84,361){\rule{0.800pt}{2.168pt}}
\multiput(593.34,361.00)(-1.000,4.500){2}{\rule{0.800pt}{1.084pt}}
\put(591.34,370){\rule{0.800pt}{1.927pt}}
\multiput(592.34,370.00)(-2.000,4.000){2}{\rule{0.800pt}{0.964pt}}
\put(588.84,378){\rule{0.800pt}{2.168pt}}
\multiput(590.34,378.00)(-3.000,4.500){2}{\rule{0.800pt}{1.084pt}}
\put(586.34,387){\rule{0.800pt}{1.927pt}}
\multiput(587.34,387.00)(-2.000,4.000){2}{\rule{0.800pt}{0.964pt}}
\multiput(585.06,395.00)(-0.560,1.096){3}{\rule{0.135pt}{1.640pt}}
\multiput(585.34,395.00)(-5.000,5.596){2}{\rule{0.800pt}{0.820pt}}
\multiput(580.06,404.00)(-0.560,0.928){3}{\rule{0.135pt}{1.480pt}}
\multiput(580.34,404.00)(-5.000,4.928){2}{\rule{0.800pt}{0.740pt}}
\multiput(575.07,412.00)(-0.536,0.797){5}{\rule{0.129pt}{1.400pt}}
\multiput(575.34,412.00)(-6.000,6.094){2}{\rule{0.800pt}{0.700pt}}
\multiput(566.43,422.40)(-0.554,0.520){9}{\rule{1.100pt}{0.125pt}}
\multiput(568.72,419.34)(-6.717,8.000){2}{\rule{0.550pt}{0.800pt}}
\multiput(557.11,430.40)(-0.611,0.516){11}{\rule{1.178pt}{0.124pt}}
\multiput(559.56,427.34)(-8.555,9.000){2}{\rule{0.589pt}{0.800pt}}
\multiput(542.70,439.40)(-1.212,0.520){9}{\rule{2.000pt}{0.125pt}}
\multiput(546.85,436.34)(-13.849,8.000){2}{\rule{1.000pt}{0.800pt}}
\put(517,445.84){\rule{3.854pt}{0.800pt}}
\multiput(525.00,444.34)(-8.000,3.000){2}{\rule{1.927pt}{0.800pt}}
\put(598.0,328.0){\rule[-0.400pt]{0.800pt}{1.927pt}}
\put(489,445.84){\rule{3.854pt}{0.800pt}}
\multiput(497.00,447.34)(-8.000,-3.000){2}{\rule{1.927pt}{0.800pt}}
\multiput(479.87,444.08)(-1.358,-0.520){9}{\rule{2.200pt}{0.125pt}}
\multiput(484.43,444.34)(-15.434,-8.000){2}{\rule{1.100pt}{0.800pt}}
\multiput(464.11,436.08)(-0.611,-0.516){11}{\rule{1.178pt}{0.124pt}}
\multiput(466.56,436.34)(-8.555,-9.000){2}{\rule{0.589pt}{0.800pt}}
\multiput(453.43,427.08)(-0.554,-0.520){9}{\rule{1.100pt}{0.125pt}}
\multiput(455.72,427.34)(-6.717,-8.000){2}{\rule{0.550pt}{0.800pt}}
\multiput(447.07,415.19)(-0.536,-0.797){5}{\rule{0.129pt}{1.400pt}}
\multiput(447.34,418.09)(-6.000,-6.094){2}{\rule{0.800pt}{0.700pt}}
\multiput(441.06,405.86)(-0.560,-0.928){3}{\rule{0.135pt}{1.480pt}}
\multiput(441.34,408.93)(-5.000,-4.928){2}{\rule{0.800pt}{0.740pt}}
\multiput(436.06,397.19)(-0.560,-1.096){3}{\rule{0.135pt}{1.640pt}}
\multiput(436.34,400.60)(-5.000,-5.596){2}{\rule{0.800pt}{0.820pt}}
\multiput(431.06,388.86)(-0.560,-0.928){3}{\rule{0.135pt}{1.480pt}}
\multiput(431.34,391.93)(-5.000,-4.928){2}{\rule{0.800pt}{0.740pt}}
\put(424.84,378){\rule{0.800pt}{2.168pt}}
\multiput(426.34,382.50)(-3.000,-4.500){2}{\rule{0.800pt}{1.084pt}}
\put(422.34,370){\rule{0.800pt}{1.927pt}}
\multiput(423.34,374.00)(-2.000,-4.000){2}{\rule{0.800pt}{0.964pt}}
\put(420.84,361){\rule{0.800pt}{2.168pt}}
\multiput(421.34,365.50)(-1.000,-4.500){2}{\rule{0.800pt}{1.084pt}}
\put(419.34,353){\rule{0.800pt}{1.927pt}}
\multiput(420.34,357.00)(-2.000,-4.000){2}{\rule{0.800pt}{0.964pt}}
\put(417.84,344){\rule{0.800pt}{2.168pt}}
\multiput(418.34,348.50)(-1.000,-4.500){2}{\rule{0.800pt}{1.084pt}}
\put(505.0,449.0){\rule[-0.400pt]{2.891pt}{0.800pt}}
\put(417.84,319){\rule{0.800pt}{2.168pt}}
\multiput(417.34,323.50)(1.000,-4.500){2}{\rule{0.800pt}{1.084pt}}
\put(419.0,328.0){\rule[-0.400pt]{0.800pt}{3.854pt}}
\put(419.34,294){\rule{0.800pt}{1.927pt}}
\multiput(418.34,298.00)(2.000,-4.000){2}{\rule{0.800pt}{0.964pt}}
\put(421.34,285){\rule{0.800pt}{2.168pt}}
\multiput(420.34,289.50)(2.000,-4.500){2}{\rule{0.800pt}{1.084pt}}
\multiput(425.38,278.86)(0.560,-0.928){3}{\rule{0.135pt}{1.480pt}}
\multiput(422.34,281.93)(5.000,-4.928){2}{\rule{0.800pt}{0.740pt}}
\multiput(430.39,271.19)(0.536,-0.797){5}{\rule{0.129pt}{1.400pt}}
\multiput(427.34,274.09)(6.000,-6.094){2}{\rule{0.800pt}{0.700pt}}
\multiput(436.40,263.37)(0.526,-0.562){7}{\rule{0.127pt}{1.114pt}}
\multiput(433.34,265.69)(7.000,-5.687){2}{\rule{0.800pt}{0.557pt}}
\multiput(443.40,254.90)(0.526,-0.650){7}{\rule{0.127pt}{1.229pt}}
\multiput(440.34,257.45)(7.000,-6.450){2}{\rule{0.800pt}{0.614pt}}
\multiput(449.00,249.08)(0.627,-0.520){9}{\rule{1.200pt}{0.125pt}}
\multiput(449.00,249.34)(7.509,-8.000){2}{\rule{0.600pt}{0.800pt}}
\multiput(459.00,241.08)(0.863,-0.516){11}{\rule{1.533pt}{0.124pt}}
\multiput(459.00,241.34)(11.817,-9.000){2}{\rule{0.767pt}{0.800pt}}
\put(474,230.34){\rule{2.600pt}{0.800pt}}
\multiput(474.00,232.34)(6.604,-4.000){2}{\rule{1.300pt}{0.800pt}}
\put(486,226.84){\rule{4.336pt}{0.800pt}}
\multiput(486.00,228.34)(9.000,-3.000){2}{\rule{2.168pt}{0.800pt}}
\put(420.0,302.0){\rule[-0.400pt]{0.800pt}{4.095pt}}
\put(504.0,227.0){\rule[-0.400pt]{1.686pt}{0.800pt}}
\put(511,160){\usebox{\plotpoint}}
\put(511,159.34){\rule{4.818pt}{0.800pt}}
\multiput(511.00,158.34)(10.000,2.000){2}{\rule{2.409pt}{0.800pt}}
\multiput(531.00,163.38)(2.271,0.560){3}{\rule{2.760pt}{0.135pt}}
\multiput(531.00,160.34)(10.271,5.000){2}{\rule{1.380pt}{0.800pt}}
\multiput(547.00,168.41)(1.020,0.507){27}{\rule{1.800pt}{0.122pt}}
\multiput(547.00,165.34)(30.264,17.000){2}{\rule{0.900pt}{0.800pt}}
\multiput(581.00,185.41)(0.556,0.507){27}{\rule{1.094pt}{0.122pt}}
\multiput(581.00,182.34)(16.729,17.000){2}{\rule{0.547pt}{0.800pt}}
\multiput(600.00,202.41)(0.527,0.507){25}{\rule{1.050pt}{0.122pt}}
\multiput(600.00,199.34)(14.821,16.000){2}{\rule{0.525pt}{0.800pt}}
\multiput(618.41,217.00)(0.509,0.657){19}{\rule{0.123pt}{1.246pt}}
\multiput(615.34,217.00)(13.000,14.414){2}{\rule{0.800pt}{0.623pt}}
\multiput(631.40,234.00)(0.514,0.877){13}{\rule{0.124pt}{1.560pt}}
\multiput(628.34,234.00)(10.000,13.762){2}{\rule{0.800pt}{0.780pt}}
\multiput(641.40,251.00)(0.516,0.990){11}{\rule{0.124pt}{1.711pt}}
\multiput(638.34,251.00)(9.000,13.449){2}{\rule{0.800pt}{0.856pt}}
\multiput(650.38,268.00)(0.560,2.439){3}{\rule{0.135pt}{2.920pt}}
\multiput(647.34,268.00)(5.000,10.939){2}{\rule{0.800pt}{1.460pt}}
\put(653.34,285){\rule{0.800pt}{4.095pt}}
\multiput(652.34,285.00)(2.000,8.500){2}{\rule{0.800pt}{2.048pt}}
\put(653.84,302){\rule{0.800pt}{8.191pt}}
\multiput(654.34,302.00)(-1.000,17.000){2}{\rule{0.800pt}{4.095pt}}
\put(652.34,353){\rule{0.800pt}{4.095pt}}
\multiput(653.34,353.00)(-2.000,8.500){2}{\rule{0.800pt}{2.048pt}}
\put(649.84,370){\rule{0.800pt}{4.095pt}}
\multiput(651.34,370.00)(-3.000,8.500){2}{\rule{0.800pt}{2.048pt}}
\put(646.34,387){\rule{0.800pt}{3.600pt}}
\multiput(648.34,387.00)(-4.000,9.528){2}{\rule{0.800pt}{1.800pt}}
\multiput(644.07,404.00)(-0.536,1.690){5}{\rule{0.129pt}{2.467pt}}
\multiput(644.34,404.00)(-6.000,11.880){2}{\rule{0.800pt}{1.233pt}}
\multiput(638.08,421.00)(-0.526,1.351){7}{\rule{0.127pt}{2.143pt}}
\multiput(638.34,421.00)(-7.000,12.552){2}{\rule{0.800pt}{1.071pt}}
\multiput(631.08,438.00)(-0.516,0.990){11}{\rule{0.124pt}{1.711pt}}
\multiput(631.34,438.00)(-9.000,13.449){2}{\rule{0.800pt}{0.856pt}}
\multiput(622.08,455.00)(-0.511,0.671){17}{\rule{0.123pt}{1.267pt}}
\multiput(622.34,455.00)(-12.000,13.371){2}{\rule{0.800pt}{0.633pt}}
\multiput(607.65,472.41)(-0.525,0.507){27}{\rule{1.047pt}{0.122pt}}
\multiput(609.83,469.34)(-15.827,17.000){2}{\rule{0.524pt}{0.800pt}}
\multiput(588.68,489.41)(-0.680,0.507){27}{\rule{1.282pt}{0.122pt}}
\multiput(591.34,486.34)(-20.338,17.000){2}{\rule{0.641pt}{0.800pt}}
\multiput(562.79,506.40)(-1.179,0.516){11}{\rule{1.978pt}{0.124pt}}
\multiput(566.90,503.34)(-15.895,9.000){2}{\rule{0.989pt}{0.800pt}}
\put(535,514.34){\rule{3.400pt}{0.800pt}}
\multiput(543.94,512.34)(-8.943,4.000){2}{\rule{1.700pt}{0.800pt}}
\put(522,517.34){\rule{3.132pt}{0.800pt}}
\multiput(528.50,516.34)(-6.500,2.000){2}{\rule{1.566pt}{0.800pt}}
\put(655.0,336.0){\rule[-0.400pt]{0.800pt}{4.095pt}}
\put(490,517.34){\rule{2.891pt}{0.800pt}}
\multiput(496.00,518.34)(-6.000,-2.000){2}{\rule{1.445pt}{0.800pt}}
\put(474,514.34){\rule{3.400pt}{0.800pt}}
\multiput(482.94,516.34)(-8.943,-4.000){2}{\rule{1.700pt}{0.800pt}}
\multiput(465.79,512.08)(-1.179,-0.516){11}{\rule{1.978pt}{0.124pt}}
\multiput(469.90,512.34)(-15.895,-9.000){2}{\rule{0.989pt}{0.800pt}}
\multiput(448.09,503.09)(-0.773,-0.507){27}{\rule{1.424pt}{0.122pt}}
\multiput(451.05,503.34)(-23.045,-17.000){2}{\rule{0.712pt}{0.800pt}}
\multiput(426.09,483.64)(-0.507,-0.527){25}{\rule{0.122pt}{1.050pt}}
\multiput(426.34,485.82)(-16.000,-14.821){2}{\rule{0.800pt}{0.525pt}}
\multiput(410.09,466.37)(-0.509,-0.569){21}{\rule{0.123pt}{1.114pt}}
\multiput(410.34,468.69)(-14.000,-13.687){2}{\rule{0.800pt}{0.557pt}}
\multiput(396.08,449.04)(-0.512,-0.788){15}{\rule{0.123pt}{1.436pt}}
\multiput(396.34,452.02)(-11.000,-14.019){2}{\rule{0.800pt}{0.718pt}}
\multiput(385.08,430.11)(-0.520,-1.139){9}{\rule{0.125pt}{1.900pt}}
\multiput(385.34,434.06)(-8.000,-13.056){2}{\rule{0.800pt}{0.950pt}}
\multiput(377.07,410.76)(-0.536,-1.690){5}{\rule{0.129pt}{2.467pt}}
\multiput(377.34,415.88)(-6.000,-11.880){2}{\rule{0.800pt}{1.233pt}}
\multiput(371.07,393.76)(-0.536,-1.690){5}{\rule{0.129pt}{2.467pt}}
\multiput(371.34,398.88)(-6.000,-11.880){2}{\rule{0.800pt}{1.233pt}}
\put(364.34,370){\rule{0.800pt}{4.095pt}}
\multiput(365.34,378.50)(-2.000,-8.500){2}{\rule{0.800pt}{2.048pt}}
\put(361.84,353){\rule{0.800pt}{4.095pt}}
\multiput(363.34,361.50)(-3.000,-8.500){2}{\rule{0.800pt}{2.048pt}}
\put(502.0,520.0){\rule[-0.400pt]{4.818pt}{0.800pt}}
\put(362.34,285){\rule{0.800pt}{7.000pt}}
\multiput(360.34,304.47)(4.000,-19.471){2}{\rule{0.800pt}{3.500pt}}
\multiput(367.38,272.88)(0.560,-2.439){3}{\rule{0.135pt}{2.920pt}}
\multiput(364.34,278.94)(5.000,-10.939){2}{\rule{0.800pt}{1.460pt}}
\multiput(372.39,257.76)(0.536,-1.690){5}{\rule{0.129pt}{2.467pt}}
\multiput(369.34,262.88)(6.000,-11.880){2}{\rule{0.800pt}{1.233pt}}
\multiput(378.40,244.52)(0.514,-0.877){13}{\rule{0.124pt}{1.560pt}}
\multiput(375.34,247.76)(10.000,-13.762){2}{\rule{0.800pt}{0.780pt}}
\multiput(388.40,228.04)(0.512,-0.788){15}{\rule{0.123pt}{1.436pt}}
\multiput(385.34,231.02)(11.000,-14.019){2}{\rule{0.800pt}{0.718pt}}
\multiput(399.41,212.37)(0.509,-0.569){21}{\rule{0.123pt}{1.114pt}}
\multiput(396.34,214.69)(14.000,-13.687){2}{\rule{0.800pt}{0.557pt}}
\multiput(412.00,199.09)(0.587,-0.507){27}{\rule{1.141pt}{0.122pt}}
\multiput(412.00,199.34)(17.631,-17.000){2}{\rule{0.571pt}{0.800pt}}
\multiput(432.00,182.09)(0.989,-0.507){27}{\rule{1.753pt}{0.122pt}}
\multiput(432.00,182.34)(29.362,-17.000){2}{\rule{0.876pt}{0.800pt}}
\multiput(465.00,165.06)(1.936,-0.560){3}{\rule{2.440pt}{0.135pt}}
\multiput(465.00,165.34)(8.936,-5.000){2}{\rule{1.220pt}{0.800pt}}
\put(479,159.34){\rule{4.818pt}{0.800pt}}
\multiput(479.00,160.34)(10.000,-2.000){2}{\rule{2.409pt}{0.800pt}}
\put(362.0,319.0){\rule[-0.400pt]{0.800pt}{8.191pt}}
\put(499.0,160.0){\rule[-0.400pt]{2.891pt}{0.800pt}}
\put(244,505){\usebox{\plotpoint}}
\multiput(244.00,503.09)(0.735,-0.500){669}{\rule{1.376pt}{0.121pt}}
\multiput(244.00,503.34)(494.143,-338.000){2}{\rule{0.688pt}{0.800pt}}
\sbox{\plotpoint}{\rule[-0.200pt]{0.400pt}{0.400pt}}%
\put(505,336){\makebox(0,0){$+$}}
\end{picture}

\end{center}

\vspace{0.3cm}
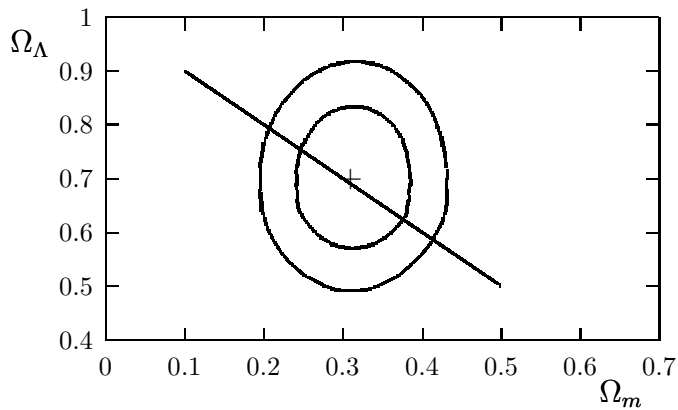
Fig. 1: The 1$\sigma$ and 2$\sigma$ statistical 
confidence regions in the ($\Omega_m,\Omega_{\Lambda}$)-plane 
are shown. The '+' marks the best fit: ($\Omega_m$,$\Omega_{\Lambda}$) 
= (0.31,0.70). The diagonal line corresponds to a flat cosmology.
\vspace{0.8cm} 

In Fig.1 we plot the shape of the $1\sigma$ and $2\sigma$ contours.
In an exactly flat cosmology the $\Omega_m$ value is the same, but the
error is only 0.04.

Since $\Omega_m + \Omega_{\Lambda}= 1.01\pm 0.15$, the important
conclusions are that (i) the data strongly prefer a flat cosmology, (ii)
the Einstein-de Sitter model is very convincingly ruled out, and (iii)
also any low-density model with $\Omega_{\Lambda}=0$ is ruled out. 

Substituting the above parameter values into the Friedman-Lemaître model
Eq.  (\ref{f3}) for the case of a flat cosmology, and taking the Hubble
constant to be known to an accuracy of 7.4\% (Nevalainen \& Roos 1998;
Giovanelli et al.  1997), yields an age estimate for the Universe of
\begin{eqnarray} 
t_0 = 13.7^{+1.2}_{-1.1}\ (0.68/h)\ \rm{Gyr}\ . 
\end{eqnarray} 
Note that this very accurate new result is purely
cosmological, completely independent of estimates of the ages of stars
or on the onset of star formation.\\

\noindent{\bf Acknowledgements:} The authors wish to thank D. Groom,
Berkeley, K.  Mattila, Helsinki, J.  Nevalainen, Harvard, M. Tegmark, 
Princeton, and O. Vilhu, Helsinki for useful
comments.  S.M.H.  is indebted to the Ella and Georg Ehrnrooth
Foundation and the E.J.  Sariola Foundation for their support.

\end{document}